\RequirePackage{fix-cm}
\documentclass[smallextended]{svjour3}       % onecolumn (second format)
\smartqed  % flush right qed marks, e.g. at end of proof
\usepackage{graphicx}
\usepackage{braket}
\usepackage{pgfplots}
\usepackage{csquotes}
\usepackage{float}
\usepackage{multirow}
\usepackage{dcolumn}
\usepackage[colorlinks=true, citecolor=blue]{hyperref}
\usepackage{hhline}
\usepackage{amssymb}
\usepackage{amsmath}
\usepackage{graphicx}
\usepackage{subfigure}
\begin{document}

\title{Non classical light in $J_x$ photonic lattice %\thanks{Grants or other notes
}


\author{Manoranjan Swain \and Amit Rai %\and
      %  Bikash K. Behera \and Prasanta K. Panigrahi %etc.
}


\institute{Manoranjan Swain \at
             Department of Physics and Astronomy,
              National Institute of Technology, Rourkela, 769008, Odisha, India \\       \email{swainmanoranjan333@gmail.com} 
             \and
              Amit Rai \at
              Department of Physics and Astronomy,
              National Institute of Technology, Rourkela, 769008, Odisha, India\\
            \email{amitrai007@gmail.com}          
}

\date{Received: date / Accepted: date}

\maketitle

\begin{abstract}
We report the study of non-classical light in a photonic lattice having parabolic coupling distribution, also known as $J_x$ photonic lattice. We focus on two photon Fock state, two photon $N$00$N$ state, and single mode squeezed state along with coherent state as input to the lattice. We investigate the possibility of perfect transfer of mean photon number as well as quantum state from one waveguide mode to other. We study photon-photon correlation for two photon $N$00$N$ state. For single mode squeezed state we study in detail, the evolution of squeezing factor and entanglement between the waveguide modes. Our findings suggest perfect transfer of average photon number for all cases and perfect transfer of quantum state in case of two photon Fock state and two photon $N$00$N$ state only and not in the case of squeezed and coherent state. Our results should have applications in physical implementation of photonic continuous variable quantum information processing.
\end{abstract}

\section{Introduction}
 Optical photonic lattices have emerged as promising platform for the study of quantum optics and quantum information as photon propagation through them ensures a very low loss~\cite{pnf_19,politi}. These coupled waveguide lattices are integrated structures that provide relatively less decoherence compared to bulk optical elements~\cite{PT1,PT2}  and are scalable. The possibility to tune parameters such as, onsite propagation constant and nearest neighbour coupling allows the control and manipulation of light propagating through them. Using uniformly coupled waveguides quantum walk~\cite{pnf_19}, entanglement generation~\cite{PT4}, quantum interference~\cite{PT5}, optical Bloch oscillation~\cite{PT6}, boson sampling~\cite{PT7} are successfully studied. In addition to uniformly coupled lattices, it is also possible to engineer lattices with non-uniform coupling such as Glauber Fock lattices~\cite{PT8} and $J_x$ photonic lattices~\cite{PT9}. These specially designed waveguide structures are useful for specific tasks. For example Glauber Fock photonic lattice with square root coupling has been used in observing interesting quantum correlation which was not found in case of uniform arrays~\cite{PT10,PT11}. Similarly $J_x$ photonic lattices with parabolic coupling can be considered as optical analogue of spin chains (as the presence or absence of photon in individual waveguides is considered as the required qubit states $|1\rangle$ and $|0\rangle$) and finds application in optical implementation of perfect state transfer (PST)~\cite{PT9}. 
 
 Achieving PST is essential for quantum computation and information processing as it aims long range communication between different parties. The idea of PST was first proposed for uniformly coupled spin-chains~\cite{PT12} which was useful for short range communication. Studies related to PST based on spin chains can be found in Ref.~\cite{new1,new2,new3,new4,new5,new6,new7,new8}. However the requirement of specially engineered nearest neighbour coupling for long range communication has become an obstacle in experimental realization and is difficult to implement with current technology. On the other hand advances in technologies allows fabrication of photonic lattices with desired coupling constant~\cite{PT1,PT2}, hence making them suitable to observe PST. Earlier theoretical studies on PST using integrated waveguide lattices can be found in Ref.~\cite{PT9,pnew2,npt22} with successful experimental implementations in Ref.~\cite{pnew3,pnew4,PT13}.

In this paper we study the transport of non-classical light~\cite{PT14} in $J_x$ photonic lattice. Recently non-classical light has received wide attention not only in quantum optics but also in quantum information processing. Our focus is on interesting non-classical states namely, two photon Fock state~\cite{2p1,new13,new14}, two photon $N$00$N$ state~\cite{PT16}, single mode squeezed state~\cite{PT17}. Note that these non-classical states find intriguing applications in quantum metrology~\cite{PT18,PT19,me1}, quantum computation and quantum information processing~\cite{PT22}, continuous variable entanglement generation~\cite{pp11,pp12}. Our aim is to address a general question, \emph{whether the perfect transfer of average photon number and quantum state is possible for these non-classical states or not?} Earlier studies on perfect state transfer has focused on perfect transfer of path-entangled states~\cite{PT9}, experimental coherent transport of light~\cite{pnew4} and experimental transfer of an entangled state~\cite{PT13}. We investigate the possibility of prefect transfer of average photon number and quantum state to another waveguide as well as the revival of both to the input waveguide. We calculate fidelity as a measure of similarity between the transferred state and the ideal state. In addition to fidelity we also study the evolution of the photon-photon correlation for two photon $N$00$N$ state. For single mode squeezed state we study the evolution of squeezing parameter and entanglement between waveguide modes. Apart from non-classical state we also study the transfer and revival of coherent state~\cite{coherent}.

Our work in this paper is organized as follows. In section \ref{model} we describe in detail about the theoretical model and the physically realizable parameters important for our study. In section \ref{ncs} we show the transport of non-classical lights in $J_x$ photonic lattice and discuss our results. We finally conclude in section \ref{con}.

\section{Waveguide model}\label{model}

\begin{figure}[h]
  \centering
  \includegraphics[width=0.8\textwidth]{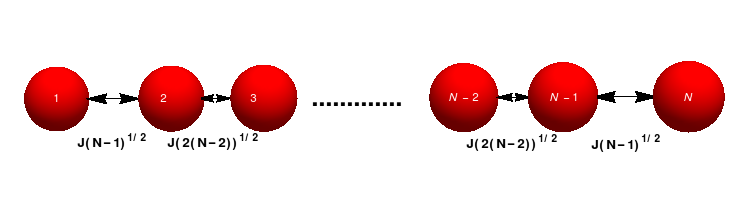}
  \caption{Schematic view of $J_x$ photonic lattice with parabolic coupling distribution.}
  \label{wp}
  \end{figure}

The interaction Hamiltonian of the system we have considered is given as, 
\begin{equation}
    \hat{H}= \sum_{j=1}^{N-1} J_j(\hat{a}_j^\dagger\hat{a}_{j+1}+\hat{a}_{j}\hat{a}^\dagger_{j+1})
\end{equation}
where $\hat{a}_j$ and $\hat{a}^\dagger_j$ are the annihilation and creation operators corresponding to $j$th mode. $J_j$ is the coupling strength between $\emph{j}$th and $(\emph{j+1}$)th guide. For perfect state transfer the coupling parameter $J_j$ is given by~\cite{PT9},
\begin{equation}
    J_j=J\sqrt{j(N-j)}
\end{equation}
where $J$ is the characteristic coupling strength and N is the number of waveguides. This type of coupling can be achieved by maintaining suitable relative separation between the individual waveguides. The waveguide structure is represented in Fig.~\ref{wp}.

The evolution of light in waveguide system can be analyzed by means of Heisenberg's equation of motion, which is given as,
\begin{equation}
    i \frac{d\hat{a}_1}{dt}=J_1\hat{a}_2
\end{equation}
\begin{equation}
    i \frac{d\hat{a}_j}{dt}=J_{j-1}\hat{a}_{j-1}+ J_j\hat{a}_{j+1}
\end{equation}
\begin{equation}
    i \frac{d\hat{a}_N}{dt}=J_{N-1}\hat{a}_{N-1}
\end{equation}
The above set of linear equations can be solved as,
\begin{equation}
   i\dot{A}_{1,l}= J_1 A_{2,l},
\end{equation}
\begin{equation}
   i\dot{A}_{j,l}= J_{j-1}A_{j-1,l}+J_j A_{j+1,l},
\end{equation}

\begin{equation}
   i\dot{A}_{N,l}= J_{N-1}A_{N-1,l},
\end{equation}
\begin{equation}
     A_{j,l}(t=0)= \delta_{jl}
\end{equation}

\begin{equation}\label{te}
     \hat{a}_j(t)=  \sum_{l}A_{j,l}(t)\hat{a}_l(0)
\end{equation}
where $\delta_{jl}$ is the Kronecker delta function. 
The average number of photons at the output of $j$th waveguide can be calculated by,
\begin{equation}\label{in}
    N_j=\langle \hat{a}_j^\dagger (t) \hat{a}_j(t)\rangle
\end{equation}
For two photon $N$00$N$ state we study the photon-photon correlation of detecting one photon at waveguide `m' and another at `n'. The photon-photon correlation function is given as~\cite{PT5},
\begin{equation}\label{cf}
    p_{n,m}=\langle \hat{a}_m^\dagger(t)\hat{a}_n^\dagger(t)\hat{a}_n(t)\hat{a}_m(t)\rangle
\end{equation}
To study the propagation of quadrature squeezed light in waveguide array we introduce the quadrature operators q$_j\equiv (\hat{a}_j+\hat{a}_j^{\dagger})/\sqrt{2}$ and p$_j\equiv(\hat{a}_j-\hat{a}_j^{\dagger})/\sqrt{2}i$ . Further the squeezing factors are given as, s$_{j}$(q) $\equiv$ ($\Delta q_{j}$)$^2$-$\frac{1}{2}$ and s$_{j}$(p)$\equiv (\Delta p_{j})^2$-$\frac{1}{2}$. These can be simplified for the state $|\psi_s\rangle = \text{exp}\big(\frac{1}{2}\xi \hat{a}_{l}^{\dagger 2}-\frac{1}{2}\xi^* \hat{a}_{l}^{2}\big)|0\rangle$ as~\cite{PT25},
\begin{equation}\label{sq}
    s_j=|A_{j,l}|^2 \text{sinh}^2r \mp \frac{1}{4} \text{sinh} 2r[A_{j,l}^2 e^{(i \phi)} + c.c.]
\end{equation}
where the -ve and +ve sign indicates quadrature p and q respectively. The negative value of these parameters will indicate squeezing in respective quadratures. The squeezing parameter can be experimentally measured by homodyne detection~\cite{nn1,nn2,PT24}. We also examine the correlation between different waveguide modes using entanglement correlation function given as~\cite{PT25,new11,new12},
\begin{equation}\label{mq}
    M(j,k)=\langle a_j^\dagger a_j\rangle + \langle a_k^\dagger a_k\rangle + \langle a_j a_k\rangle + \langle a_j^\dagger a_k^\dagger\rangle
\end{equation}
The negative value of M(j,k) is both necessary and sufficient condition for verification of entanglement for Gaussian states.

\section{Non-classical light as input to the waveguide}\label{ncs}
\subsection{\textbf{Two photon Fock state}}
Let's consider that the two photon Fock state is fed to the $l$th waveguide.
\begin{figure}[h]
  \centering
  \includegraphics[height=0.3\textwidth,width=0.8\textwidth]{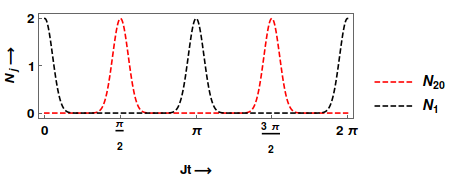}
  \caption{Mean photon number at 1$^{st}$ and 20$^{th}$ waveguide with respect to time when two photons are injected to the 1$^{st}$ waveguide.}
  \label{2pin}
\end{figure}\\
 The input state can be written as, 
\begin{equation}
    |\psi\rangle =\frac{1}{\sqrt{2}} \hat{a}_l^{\dagger 2}|0\rangle
\end{equation}
The dynamics of the state evolution can be studied by solving Heisenberg's equation of motion. For the two photon Fock state Eq. $ \ref{in}$ can be given as, $  N_j=2|A_{jl}(t)|^2$. $N_j$ represents the average photon number at the $jth$ waveguide. The evolution of mean photon number for two photon input state with respect to time is shown in Fig. \ref{2pin}. From Fig.~\ref{2pin} it can be noticed that the complete transfer of average photon number takes place from 1$^{st}$ waveguide to the 20$^{th}$ waveguide at Jt=$\pi/2$. We also find the complete revival of the average photon number at the input waveguide at Jt=$\pi$. Now to verify the process of perfect state transfer we have calculated fidelity. The fidelity function is defined as,
\begin{equation}\label{fidelity}
    F= |\langle \Psi|\psi(t)\rangle|^2
\end{equation}
where $|\Psi\rangle$ is the state which one should expect at the time of perfect state transfer and $|\psi(t)\rangle$ is the evolved state of the input given in Eq. $ \ref{in}$. In the present case $|\Psi\rangle$ can be given as,
\begin{equation}
    |\Psi\rangle=\frac{1}{\sqrt{2}} \hat{a}_{20}^{\dagger 2}(0)|0\rangle
\end{equation}

%The state at Jt=$\pi/2$ can be written as, 
%\begin{equation}
   % |\psi(\pi/2)\rangle= \frac{1}{\sqrt{2}}(A_{20,1}^*(\pi/2)\hat{a}_{20}^\dagger(0))^2|0\rangle
%\end{equation}
For the two photon input, fidelity is calculated to be 1 at Jt=$\pi/2$. This shows the complete transfer of two photon state from first waveguide to the last waveguide.

\subsection{\textbf{Two photon $N$00$N$ state}}
\begin{figure}[h]
  \centering
  \includegraphics[height=0.3\textwidth,width=0.8\textwidth]{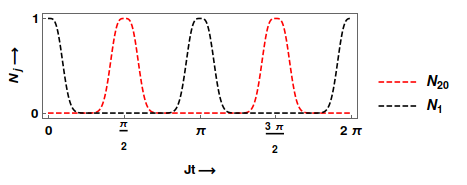}
  \caption{Mean photon number plot for two photon $N$00$N$ state for p=1 and q=2. The plots for N$_{2}$ and N$_{19}$ are similar as that of N$_{1}$ and N$_{20}$.}
  \label{$N$OO$N$intensity}
  \end{figure}

$N$00$N$ states hold important applications in quantum metrology and quantum sensing as they can be used for precise phase measurements~\cite{me1}. The initial state of the system with the two photon $N$00$N$ state as input can be given as,
\begin{equation}
    |\psi_N\rangle =\frac{1}{2} (\hat{a}_p^{\dagger 2}+\hat{a}_q^{\dagger2})|00\rangle
\end{equation}
where p and q are the input sites. The evolution of $N$00$N$ state in waveguide can be studied in the similar manner as discussed for two photon Fock state. The average photon number at the $j$th waveguide is given by, $N_j=|A_{j,p}(t)|^2+|A_{j,q}(t)|^2$. The time evolution of the mean photon number is shown in Fig.~\ref{$N$OO$N$intensity}. The plot shows the complete transfer of average photon number to 19$^{th}$ and 20$^{th}$ waveguide (input is given at first two waveguide) at Jt=$\pi/2$ and revival of the same to the input waveguides at Jt=$\pi$.

\begin{figure*}[h]
\centering
\subfigure[]{\includegraphics[width=0.4\linewidth]{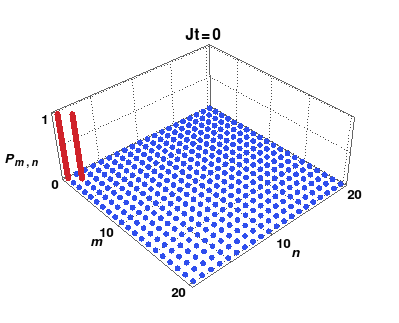}}\quad
\subfigure[]{ \includegraphics[width=0.4\linewidth]{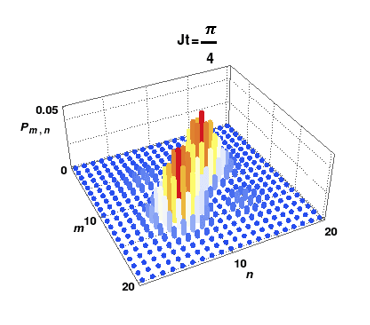}}\quad
\subfigure[]{ \includegraphics[width=0.4\linewidth]{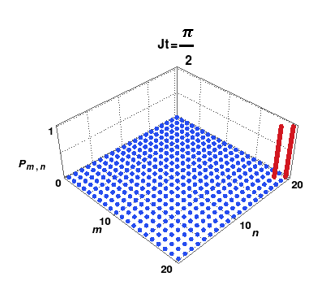}}\quad
\caption{Photon-photon correlation function at different times for two-photon $N$00$N$ state.}
 \label{ncorrelation}
\end{figure*}

\noindent To verify the process of perfect state transfer, we calculate fidelity in similar manner as discussed in case of two photon Fock state. The expected state at perfect transfer is given as,
\begin{equation}
    |\Psi_N\rangle =\frac{1}{2} (\hat{a}_{19}^{\dagger 2}(0)+\hat{a}_{20}^{\dagger2}(0))|00\rangle
\end{equation}
%The state at Jt=$\pi/2$ can be given as,
%\begin{equation}
  %  |\psi_N\rangle =\frac{1}{\sqrt{2}} (|A_{20,1}^*(\pi/2)|a_{20}^{\dagger 2}+|A_{19,2}^*(\pi/2)|a_{19}^{\dagger2})|00\rangle
%\end{equation}
Using Eq. $\ref{fidelity}$ the fidelity of $N$00$N$ state transfer is calculated to be 1. Next we study the photon-photon correlation by means of photon-photon correlation function. The photon-photon correlation function for the two photon $N$00$N$ state for getting one photon at m$^{th}$ and other at n$^{th}$ guide is calculated as, $p_{n,m}= |A_{n,1}A_{m,1}+A_{n,2}A_{m,2}|^2$. The plots of photon-photon correlation function at different times is shown in Fig.~\ref{ncorrelation}. The plots show perfect transfer of state at Jt=$\pi/2$. Fig.~\ref{ncorrelation}(b) represents the antibunching effect of two photons during their propagation in waveguide lattice. 
  
\noindent

\subsection{\textbf{Single mode squeezed state}}
Now we consider single mode squeezed state as an input to the coupled waveguide system. Squeezed states are useful for high-precision optical measurements~\cite{se1}, radiometry~\cite{se2}, quantum information processing~\cite{PT22}. With this squeezed state input, the initial state of the system is given as, 
\begin{equation}
   |\psi_s\rangle = \text{exp}\big(\frac{1}{2}\xi \hat{a}_{l}^{\dagger 2}-\frac{1}{2}\xi^* \hat{a}_{l}^{2}\big)|0\rangle
\end{equation}
where l is input site, $\xi$= r e$^{i \phi}$, r is the squeezing parameter and $\phi$ represents squeezing direction.
%The state after time t is given as, 

%\begin{equation}
%  |\psi(t)\rangle = exp(\frac{1}{2}\xi a_{l}^{\dagger 2}(-t)-\frac{1}{2}\xi^* a_{l} ^{2}(-t))
%\end{equation}
Using Eq.  $ \ref{in}$ the average photon number at waveguides for this case can be written as,
\begin{equation}
    N_j=\text{sinh}^2r |A_{jl}|^2
\end{equation}

\begin{figure}[h]
  \centering
  \subfigure[]{\includegraphics[height=0.3\textwidth,width=0.8\textwidth]{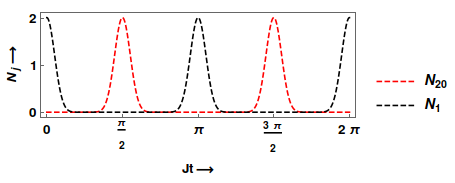}}\quad
\subfigure[]{ \includegraphics[height=0.3\textwidth,width=0.85\textwidth]{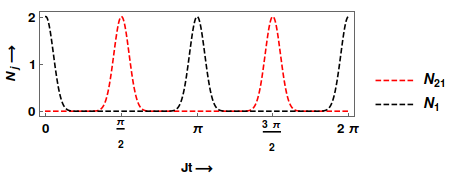}}\quad
\caption{Average photon number variation at input and output waveguide for (a) N=20 and (b) N=21. The value of r is set to be 1.15.}
  \label{squeezingintensity}
  \end{figure}
\begin{figure*}[h]
\centering
\subfigure[]{\includegraphics[width=0.4\linewidth]{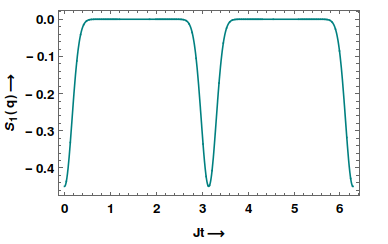}}\quad
\subfigure[]{ \includegraphics[width=0.4\linewidth]{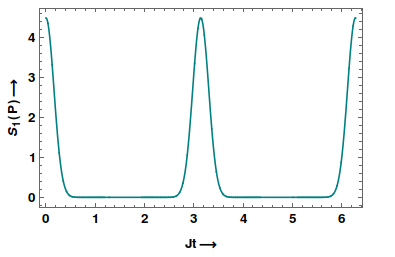}}\quad
\subfigure[]{ \includegraphics[width=0.4\linewidth]{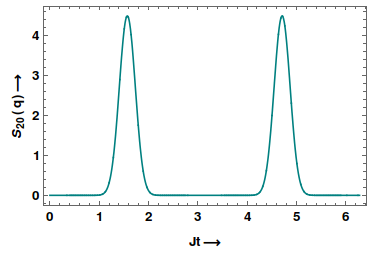}}\quad
\subfigure[]{ \includegraphics[width=0.4\linewidth]{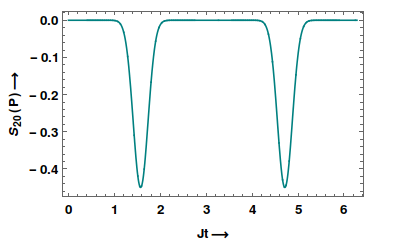}}\quad
\subfigure[]{\includegraphics[width=0.4\linewidth]{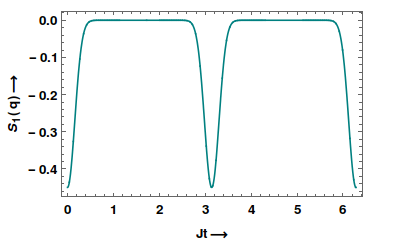}}\quad
\subfigure[]{ \includegraphics[width=0.4\linewidth]{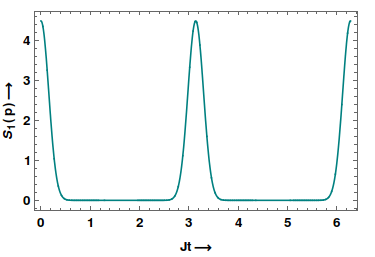}}\quad
\subfigure[]{ \includegraphics[width=0.4\linewidth]{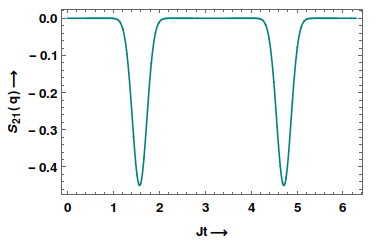}}\quad
\subfigure[]{ \includegraphics[width=0.4\linewidth]{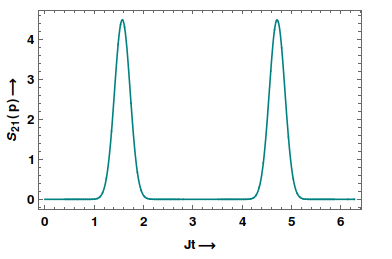}}\quad
 \caption{Initial squeezing factors (a) and (b) for N=20 (e) and (f) for N=21 and final squeezing factors (c) and (d) for N=20 and (g) and (h) for N=21 . The value of r is set to be 1.15. $\phi$ is considered to be $\pi$.}
 \label{squeezingparameters}
\end{figure*}

\noindent 
We consider two photonic lattices with 20 and 21 coupled waveguides respectively. In both the platforms we found the complete transfer of average photon number. Fig.~\ref{squeezingintensity} shows the complete transfer of average photon number of squeezed light from the first waveguide to the 20$^{th}$ waveguide (for N=20) and 21$^{st}$ waveguide (for N=21) at Jt=$\pi/2$ respectively. For odd number of waveguides (i.e. N=21) the fidelity is calculated to be one. For even number of waveguides (i.e. N=20) the fidelity is calculated to be $\frac{1}{\text{cosh}(2r)}$. This suggests that, the state one should expect at Jt=$\pi/2$ at the output waveguide when even number of waveguides are considered is not exactly same as the input state, although the average photon number of light is found to be same in both. This can be explained on the basis of evolution of squeezing factors, which can be done using Eq. $\ref{sq}$. Fig.~\ref{squeezingparameters} depicts the evolution of squeezing factors for the first and the last guide with time.\\
\\
From Fig.~\ref{squeezingparameters} it can be noticed that in case of 20 (even) waveguides (Fig.~\ref{squeezingparameters}. a-d) initially the q quadrature of the input state is squeezed however at Jt=$\pi/2$ the squeezing is found to be in the p quadrature and not in the q quadrature. For the case of 21 (odd) waveguides (Fig.~\ref{squeezingparameters}. e-f) the squeezing is found in q quadrature both at Jt=0 and at Jt=$\pi/2$. Hence the perfect state transfer also depends on the number of waveguides that are present in the lattice. Note that the fidelity of state transfer in the latter case further decreases with increase in r. For squeezed state input we observed the perfect revival of state to the input guide at Jt=$\pi$. 

Our results can be explained by studying the evolution of the initial state as it propagates through the $J_x$ photonic lattices. We consider the initial state of the system can be given as,
\begin{equation}
   |\psi(0)\rangle = \sum_n C_n|n\rangle
\end{equation}

The state represents wide class of quantum states. By choosing appropriate values of $C_n$ different states can be realized. For example the squeezed state can be represented as,
\begin{equation}
\begin{aligned}
    |\psi_s(0)\rangle = \frac{1}{\sqrt{\text{cosh}r}} \text{exp}\bigg(e^{i\phi} \text{tanh}r \frac{\hat{a}_{1}^{\dagger 2}}{2}\bigg)|0\rangle\\
    =\frac{1}{\sqrt{\text{cosh}r}}\sum_{n=0}^\infty \text{e}^{i n \phi}\big(\text{tanh}r\big)^n \frac{\sqrt{(2n)!}}{n!2^n}|2n\rangle
\end{aligned}
\end{equation}

The evolution of the creation and the annihilation operators can be studied in Heisenberg picture. For 20 waveguide the annihilation operator at t=0 i.e. $\hat{a}_1$ evolves to i$\hat{a}_{20}$ at Jt=$\pi/2$. Hence the state at Jt=$\pi/2$ can be given as,

\begin{equation}\label{25}
\begin{aligned}
     |\psi_s(\pi/2J)\rangle = \frac{1}{\sqrt{\text{cosh}r}} \text{exp}\bigg(e^{i\phi} \text{tanh}r \frac{(i\hat{a}_{20})^{\dagger 2}}{2}\bigg)|0\rangle
\\= \text{exp}\bigg(-\frac{1}{2}\xi \hat{a}_{20}^{\dagger 2}+\frac{1}{2}\xi^* \hat{a}_{20}^{2}\bigg)|0\rangle
\end{aligned}
\end{equation}

whereas, the initial state of the system was,
\begin{equation}\label{26}
   |\psi_s(0)\rangle = \text{exp}\bigg(\frac{1}{2}\xi \hat{a}_{1}^{\dagger 2}-\frac{1}{2}\xi^* \hat{a}_{1}^{2}\bigg)|0\rangle
\end{equation}

It is clear from Eq. \eqref{25} and Eq. \eqref{26} that the initial state at t=0 and the state at Jt=$\pi/2$ are different. Hence the state one should expect at the time of perfect state transfer is not the same as the initial state. Further it is clear that the squeezing in the initial state is in q quadrature as shown in Fig.~\ref{squeezingparameters} (a-b) and the squeezing in the state at Jt=$\pi/2$ is in p quadrature, as shown in Fig.~\ref{squeezingparameters} (c-d). In case of a single photon
state $C_{n,m}$= $\delta_{nm}$, a global phase is introduced in the state at Jt=$\pi/2$ so in this case a prefect state transfer is observed. Similar analysis for N=21 waveguides shows that perfect state transfer will take place in this case. This is also clear from Fig.~\ref{squeezingparameters} (e-h).

\begin{figure}[h]
  \centering
  \subfigure[]{ \includegraphics[width=0.4\linewidth]{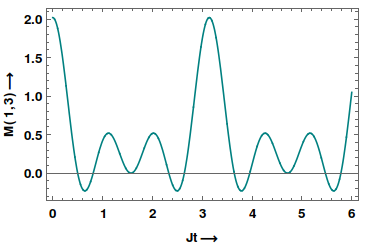}}
  \subfigure[]{ \includegraphics[width=0.42\linewidth]{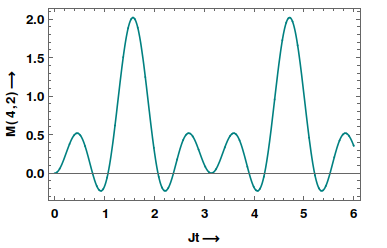}}
  \caption{(a)Variation of correlation function (a) M(1,3) for $\phi=0$. (b) M(4,2) for $\phi=\pi$. The value of r is set to be 1.15.}
  \label{m13}
  \end{figure}

We next study the entanglement between the waveguide. The presence of entanglement between waveguide modes during the propagation of squeezed light can be verified using Eq. $\ref{mq}$. For the case of four coupled waveguides,
\begin{equation}
    M(1,3)=\text{cos}^2(t) \text{sinh}(r)(-2\sqrt{3}
    \text{cos}^2(t)
    \text{cos}(\phi) \text{cosh}(r) \text{sin}^2(t)+(3 \text{sin}^4(t) + \text{cos}^4(t)) \text{sinh}(r))
\end{equation}
and 
\begin{equation}
    M(4,2)=\text{sin}^2(t)\text{sinh}(r)(2\sqrt{3}\text{cos}^2(t)\text{cos}(\phi)\text{cosh}(r)\text{sin}^2(t)+(3\text{cos}^4(t)+\text{sin}^4(t))\text{sinh}(r))
\end{equation}
The variation of M(1,3) and M(4,2) with time is shown in Fig.$~\ref{m13}$. The negative values clearly show the entanglement between the waveguide modes.
  
\subsection{\textbf{Coherent state}}
\begin{figure}[h]
  \centering
  \includegraphics[height=0.3\textwidth,width=0.8\textwidth]{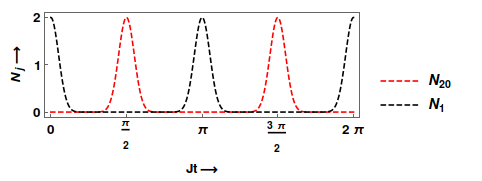}
  \caption{Mean photon number variation for coherent state when input is given at first waveguide. $|\alpha|^2$ is set to be 2.}
  \label{coherent}
  \end{figure}

We next consider a coherent state input to the waveguide, which is of the form,
\begin{equation}
    |\psi_\alpha\rangle=e^{\alpha \hat{a}_l^{\dagger}-\alpha^*\hat{a}_l} |0\rangle
\end{equation}
where $\alpha$ is a complex number and  $\hat{a}_l$ and $\hat{a}_l^\dagger$ are photonic annihilation and creation operators of the l$th$ waveguide. The average photon number at the output of j$th$ waveguide is given as, N$_j$= $|\alpha|^2 |A_{jl}|^2$. A plot for variation of N$_j$ with time is shown in Fig. \ref{coherent}.

Fig. \ref{coherent} shows the complete transfer of average photon number from 1$^{st}$ waveguide to the 20$^{th}$ waveguide at Jt=$\frac{\pi}{2}$. However the perfect state transfer results are quite different compared to all cases considered till now. We consider three photonic lattices with waveguide numbers 19, 20 and 21 respectively. For the case of 21 waveguides the fidelity of state transfer was calculated to be 1. For the case of 19 and 20 waveguides the fidelity of state transfer was calculated to be exp(-4$|\alpha|^2$) and exp(-2$|\alpha|^2$) respectively. This can be explained by the evolution of the initial state as has been done in the case of squeezed state. The coherent state can be represented in Fock basis as,
\begin{equation}
\begin{aligned}
    |\psi_{\alpha}(0)\rangle= \text{e}^{-\frac{1}{2}|\alpha|^2} \sum_{n=0}^\infty \frac{\alpha ^n}{n!}(\hat{a}_{1}^{\dagger})^n|0\rangle\\
    =\text{e}^{-\frac{1}{2}|\alpha|^2} \sum_{n=0}^\infty \frac{\alpha^n}{\sqrt{n!}}|n\rangle
\end{aligned}
\end{equation}

For N=20 waveguides the evolved state at Jt=$\pi/2$ is given as,
\begin{equation}
\begin{aligned}
    |\psi_{\alpha}(\pi/2J)\rangle= \text{e}^{-\frac{1}{2}|\alpha|^2} \sum_{n=0}^\infty \frac{\alpha ^n}{n!}((i\hat{a}_{20})^{\dagger})^n|0\rangle
   \\= e^{-i\alpha \hat{a}_{20}^{\dagger}-i\alpha^*\hat{a}_{20}} |0\rangle
\end{aligned}
\end{equation}

Due to the additional phase factor perfect state transfer is not observed in this case also. For all the lattices perfect transfer of average photon number is observed at Jt=$\frac{\pi}{2}$. We calculate the revival fidelity of coherent state to the input waveguide. The perfect revival of coherent state is observed at Jt=$\pi$ and Jt=2$\pi$ for odd and even number of guides respectively.

\section{Conclusion}\label{con}
In conclusion we analyzed the possibility PST of quantum states  in $J_x$ photonic lattice. We considered the transport of three non-classical states along with a coherent state. In all cases the complete transfer of average photon number was observed. However the genuine quantum state transfer was observed only for two photon Fock state and two photon $N$00$N$ state but not for the squeezed state and coherent state. We also studied the revival of quantum states to the input site. For all the states we have considered, we observed perfect revival of average photon number. Our findings should have important applications in quantum information processing. The possibility of PST in J$_x$ photonic lattice can further be explored for other class of non-classical states. One can study the effect of loss in photon propagation in PST. For study of photon loss Ref.~\cite{mm11,mm12,mm13} can be followed.
%\section*{Acknowledgments}

\begin{acknowledgements}
A.R. gratefully acknowledges a research grant from Science and Engineering Research Board (SERB), Department of Science and Technology (DST), Government of India (Grant No. CRG/2019/005749) during this work.
\end{acknowledgements}




\end{document}